\documentclass[acmsmall]{acmart}
\usepackage{tabularx}
\AtBeginDocument{%
  }

\setcopyright{acmlicensed}
\acmJournal{PACMHCI}
\acmYear{2025} \acmVolume{9} \acmNumber{2} \acmArticle{CSCW004} \acmMonth{4}\acmDOI{10.1145/3710902}

\begin{document}

\title{“It’s Always a Losing Game”: How Workers Understand and Resist Surveillance Technologies on the Job}

\author{Cella M. Sum}
\email{csum@andrew.cmu.edu}
\affiliation{%
  \institution{Carnegie Mellon University}
  \city{Pittsburgh}
  \state{PA}
  \country{USA}
}

\author{Caroline Shi}
\email{cshi3@andrew.cmu.edu}
\affiliation{%
  \institution{Carnegie Mellon University}
  \city{Pittsburgh}
  \state{PA}
  \country{USA}
}

\author{Sarah E. Fox}
\email{sarahf@andrew.cmu.edu}
\affiliation{%
  \institution{Carnegie Mellon University}
  \city{Pittsburgh}
  \state{PA}
  \country{USA}
}

\renewcommand{\shortauthors}{Cella M. Sum, Caroline Shi, and Sarah E. Fox}

\begin{abstract}
   With the rise of remote work, a range of surveillance technologies are increasingly being used by business owners to track and monitor employees, raising concerns about worker rights and privacy. Through analysis of Reddit posts and in-depth semi-structured interviews, this paper seeks to understand how workers across a range of sectors make sense of and respond to layered forms of surveillance. While workers express concern about risks to their health, safety, and privacy, they also face a lack of transparency and autonomy around the use of these systems. In response, workers take up tactics of everyday resistance, such as commiserating with other workers or employing technological hacks. Although these tactics demonstrate workers’ ingenuity, they also show the limitations of existing approaches to protect workers against intrusive workplace monitoring. We argue that there is an opportunity for CSCW researchers to support these countermeasures through worker-led design and policy.
\end{abstract}



\received{July 2024}
\received[revised]{October 2024}
\received[accepted]{December 2024}

\maketitle

\section{Introduction}

\begin{quote}
    “\textit{What do warehouse workers, delivery workers, and call center workers have in common? Their productivity is tracked and compared to the average of other workers. When workers are slower than average, they face disciplinary action. This causes workers to sacrifice work quality and safety to improve their times. When everyone does this, they compete against more unrealistic expectations. It’s always a losing game. Anyone can make a tracking app, but in the end it’s poor management because it doesn’t take into account worker burnout and safety. This harms both worker and employer.}” --- Post on Reddit [R141]\footnote{Throughout this paper, Reddit posts have been paraphrased or slightly altered to protect anonymity.}
\end{quote}

As the above Reddit post articulates, the current ubiquity of electronic monitoring, or Workplace Surveillance Technologies (WSTs), has transformed how workers are managed across industries. Often associated with blue-collar jobs in manufacturing and call centers since the 1990s, WSTs underwent a significant expansion in use during the COVID-19 pandemic \cite{kantor_2022}. This rise has seen these technologies flourish in previously unmonitored sectors, including higher-paying creative and technical professions. This shift has also ushered in an era of granular, real-time scrutiny for workers. A combination of “old” technologies like cameras and “new” technologies like artificial intelligence (AI) is now being deployed to track employee productivity, location, physical environment, biometrics, computer activity, and even personal data \cite{hickok2023policy}.

Paralleling this rise, there has been an increasing interest within CSCW and related scholarly communities in how data are collected about workers' performance, as well as how these evaluations affect pay structures and one’s ability to sustain employment \cite{jarrahi_2021,lee_2015}. Although initially deployed within the context of gig work, scholars have charted how algorithmic management techniques are now prevalent across traditional workplace settings and employment relationships \cite{kellogg_2020}. Research has also examined how these data-driven management strategies are combined with other forms of monitoring (e.g., customer ratings) to generate an ever more detailed assessment of workers’ activities \cite{cameron_2022,rosenblat_2016}. Scholars such as Sannon et al. \cite{sannon_2022} have documented how workers engage in risk mitigation strategies associated with such oversight, including self-protective surveillance behaviors, while others describe the potential for collective sensemaking as a strategy for worker contestation \cite{holten_moller_2021}. 

Building on this body of scholarship, our paper seeks to understand this growing trend of WSTs cross-sectorally, including impacts on those previously spared from such intrusive monitoring. In particular, we examine the following questions: (1) What are workers' lived experiences with WSTs across sectors? (2) What tactics have workers developed to circumvent harmful forms of surveillance? (3) How do workers create avenues of peer support and information sharing around the use (and circumvention) of these technologies? To do so, we draw together content analysis of a set of 9 work-related subreddits and 10 in-depth, semi-structured interviews with employees and managers from a diverse set of industries who have experienced a variety of surveillance technologies. This includes those who have worked in operations, customer service, marketing, and food and beverage service.

Workers across our study describe how WSTs contribute to a culture of distrust in the workplace that harms both workers and managers alike. Beyond privacy concerns, workers identify these technologies as causing significant stress, reducing their productivity, and increasing their risk of disciplinary action. At the same time, they see these technologies as fostering paranoia and distrust among managers toward their employees. In response, some workers employ various resistance tactics, such as commiseration, obfuscation, soldiering, and quitting to help protect them from intrusive monitoring. However, the reliance on these more individualized tactics points to the limited power and protections workers have when subjected to highly-surveilled and fissured workplaces \cite{weil2014fissured}.

Our paper makes three contributions to the CSCW community. First, we offer an empirical study on how workers have been impacted by WSTs since COVID-19 and the rise of remote work across different sectors. Second, we broaden understanding of worker resistance by highlighting the myriad tactics workers employ to navigate and circumvent surveillance and its relationship to building worker power. Third, we identify opportunities for CSCW researchers to support workers in these countermeasures and to help bridge the gap toward more collective action.

In what follows, we first describe the contemporary landscape of WSTs and their impacts on workers. We then discuss current understandings of worker resistance and how they build power in response to overt surveillance and poor working conditions, and the ways past HCI research has supported these efforts. Next, we detail the methods that guide our content analysis and interview process. Drawing on our findings that explore how workers experience and resist surveillance technologies, we demonstrate the need for HCI research to go beyond a focus on mitigating the risks of surveillance and to support efforts in resistance and refusal to its use.

\section{Related Work}
\subsection{Workplace Surveillance Technologies}
Technological advancements since the late twentieth century have led to the age of digital surveillance, making it easier for employers to constantly monitor their workers both inside and outside the workplace \cite{introna_2000,vitak_2023,ajunwa_2017}. WSTs have the ability to track and even infer a worker’s productivity, behavior, and personal information through digital activity monitoring, camera and audio monitoring, and localization and biophysical monitoring \cite{ajunwa_2017,ball_2010,cousineau_2023}. As AI and sensing technologies become more affordable and accessible, workplaces have introduced more intrusive forms of surveillance that often involve an assemblage of networked devices to generate continuous fine-grained data about their workers \cite{marx_2002}. The responsibility of workplace monitoring has also become more distributed in nature, where employees are expected to aid in their own surveillance through participation in workplace wellness programs and the use of productivity apps marketed towards improving their work and wellbeing \cite{ajunwa_2017,ball_2010,chowdhary_2023,kawakami_2023,adler_2022,saha_2023,Ajunwa_Crawford_Ford_2016}. The increased gigification of work, coupled with the rise of remote work during the COVID-19 pandemic, has seen a proliferation of these technologies being used beyond the workplace, further blurring the lines between work and private life \cite{calacci_2022,allyn_2020,satariano_2020,zickuhr_2021,gould_2023}.

As a result, there has been growing interest in HCI and adjacent fields to understand how surveillance technologies are implemented across various workplace settings \cite{mashhadi_2016}, including transportation \cite{pritchard_2015,levy_2015}, warehouses \cite{litwin_2022,cheon_2023,alimahomed-wilson_2021}, construction \cite{Kristiansen_2018}, gig work \cite{sannon_2022,anjali_anwar_2021,lee_2015}, call centers \cite{poster_2018,ball_2002,Rivera-Pelayo_2017}, domestic work \cite{ahmed_2022,johnson_2020,bernd_2020,ming_2023}, education \cite{lu_2021}, and remote work \cite{vitak_jcmc_2023}. Researchers have cited several potential benefits of its use, including the ability to offer real-time feedback for improved performance \cite{pritchard_2015,ball_2010}, mitigation of risk \cite{ball_2010}, and increased accountability \cite{barbaro_2022}.

Although these proposed benefits highlight the potential for monitoring technologies to support workers, there have been numerous concerns that undermine this vision, including threats to worker privacy, autonomy, health, and safety, and the reinforcement of existing biases \cite{awumey_2024, constantinides_2022}. These technologies often collect extraneous and irrelevant information that do not meaningfully correlate to attributes of success on the job \cite{roemmich_2023,pritchard_2015}. In particular, AI systems that capture and infer worker emotion, mood, and affect have been susceptible to inaccuracies and bias \cite{roemmich_2023,boyd_2023,corvite_2023,das_swain_2023,kaur_2022}. Workers have reported experiencing notable shifts in behavior and interpersonal dynamics, including increased stress, disciplinary action, distrust, and competition \cite{pritchard_2015,levy_2015,alimahomed-wilson_2021}. Pritchard et al. \cite{pritchard_2015} found that London bus drivers performed a form of “self-surveillance” that involved proactively altering their behavior to avoid being flagged by monitoring systems or to beat their previous performance scores. Power differentials and a lack of transparency make it difficult for workers to meaningfully consent to these technologies, as they have little power to opt out or be in a position to understand and control the extent of monitoring, their data, or how they are being assessed \cite{chowdhary_2023,sannon_2022,ajunwa_2020,Ajunwa_Crawford_Ford_2016}. In some cases, monitoring systems have led to unsafe working conditions and poorer performance, due to workers being compelled to perform additional labor in the form of continuously checking, negotiating, and responding to these systems \cite{pritchard_2015,lu_2021,sannon_2022}. Women and minority workers, in particular, have been disproportionately impacted as they are more likely to be employed in low-wage jobs and face biases that make them more susceptible to invasive monitoring \cite{ball_2010,stark_2020,anjali_anwar_2021,negron_2023}.

While there have been some studies that focus on managerial perspectives on surveillance \cite{kawakami_2023}, there has been less research on how managers experience being both the surveillant and the surveilled. In an adjacent context, Lu et al. \cite{lu_2021} found that while teachers used surveillance technologies on their students, these teachers were also subjected to monitoring by their supervisors and parents. This informed how they conducted their own surveillance, purposefully choosing to highlight or omit certain data when the act of surveillance itself became a tangible metric that reflected their job performance. Rather than viewing surveillance as a simple interaction between two actors, Lu et al. recognize this phenomenon as a more complex and “wider socio-technical assemblage” involving a myriad of different actors with conflicting expectations, incentives, and power. This may include customers, third-party platforms, and data brokers \cite{sannon_2022}, and even family members \cite{anjali_anwar_2021}.

Workers have limited legal protections in the face of workplace surveillance \cite{zickuhr_2021}. In the US, there are currently no federal laws in place to protect against invasive workplace monitoring and legislation is limited in piecemeal form at the local and state levels \cite{ajunwa_2017}. The California Privacy Rights Act of 2020 is one of the few examples that offers some privacy protections to workers \cite{gray_2020}. Although frequently used as an exemplar for omnibus legislation that provides robust data protections, there remain gaps within the European Union’s General Data Protection Regulation (GDPR) that enables organizations to surveil workers while still being in compliance \cite{jodka_2018,horan_2018}. As a result, there have been calls for stronger organizational and governmental regulation to protect worker autonomy, privacy, and collective rights \cite{das_swain_2023,ebert_2021,nguyen_2021, calacci_stein_2023} and for technologists to take a worker-centered design approach that meaningfully engages with workers as collaborators, limits unnecessary data capture, and recognizes the nuanced and contextual aspects of their labor that are harder to measure or make visible \cite{calacci_2022,holten_moller_2021,zhang_2022,kawakami_2023,cameron_2021,akridge_2024}. Our work expands upon this literature by taking a holistic view on how workers experience and make sense of surveillance technologies across sectors, particularly amid the rise of remote work during the COVID-19 pandemic and shortly thereafter. We do so by drawing from worker testimonials on Reddit and in-depth interviews.

\subsection{Worker Power and Resistance}
Since the beginning of workplace surveillance, there has been worker resistance \cite{sale_1995,luff_2008}. Earlier studies of WSTs have shown how call center workers manipulate their metrics, such as leaving lines open after hanging up or pretending to talk on the phone \cite{frenckel_1998,ball_2010,callaghan_2002}. Despite technologies becoming more sophisticated, workers have still found creative ways to circumvent them. For example, gig workers use sousveillance \cite{mann_2013,mann_2002,cecchinato_2021} techniques to monitor “up” by tracking payment discrepancies in ridehail work or collecting information about specific requesters on Amazon Mechanical Turk \cite{griffith_2022,hara_2018,do_2024,sannon_2022}. Additionally, workers use self-surveillance techniques by tracking their own personal data as a way to hold their employers and customers accountable \cite{sannon_2022,gallagher_2023}. “Worker inquiry” through data collection and analysis has long been an advocacy tool used by workers and unions since the days of Taylorism \cite{sannon_2022,Khovanskaya_2019,uermwa_1943,Khovanskaya_data_2019,dalal_2024,alquiati_2019}. Sannon et al. \cite{sannon_2022} broadly call these tactics self-protective surveillance that help to shield workers from risky work situations. They outline other resistance tactics gig workers used to protect their work and privacy, including using platforms like Reddit to commiserate with other workers, creating false personas, or using technologies like VPNs to protect their personal information \cite{sannon_2022}. Surveillance studies scholar Gary Marx outlines 11 resistance tactics in total that workers use to subvert surveillance, including discovery, avoidance, piggybacking, switching, distorting, blocking, masking, breaking, refusal, explaining and contesting, and cooperation \cite{marx_2003}. Other workers have collectively organized to form unions or participate in strikes to fight against overt surveillance and poor working conditions \cite{alimahomed-wilson_2021}.

However, each tactic of resistance carries with it specific risks and barriers that workers must negotiate \cite{do_2024}. Strategies like self-surveillance or employing technological hacks may involve significant amounts of skill and labor that workers may not find worthwhile \cite{sannon_2022}. Workers also risk facing disciplinary action or getting fired if they are caught enacting resistance. For example, gig workers have to negotiate the potential of violating platform policies or damaging requester relationships when practicing sousveillance due to the potential risk of getting kicked off the platform \cite{do_2024}. Sannon and colleagues highlight that resistance under surveillance serves as a reflection and a response to the unequal power relations that exist between workers and their surveillant actors \cite{sannon_2022}. The presence of surveillance technologies itself has been used as a deterrent to building worker power and unionization \cite{zickuhr_2021,alimahomed-wilson_2021}.

To address these power asymmetries, some HCI researchers have investigated how technological tools can better support workers in these countermeasures \cite{hara_2018,salehi_2015,irani_2013,calacci_pentland_2022,zhang_2023,toxtli_2023,toxtli_2021,georgia_2024,driversseat,timeproject,lee_2015}. One of the most well-known examples is Turkopticon, an online platform where Amazon Mechanical Turk workers can rate, review, and vet requestors. Calacci \& Pentland \cite{calacci_pentland_2022} co-built the Shipt calculator with worker groups to allow gig workers to track and share wage data and surface pay disparities. While these two tools have helped support workers in collective data stewardship, sensemaking, and organizing, Calacci states that more can be done to support “Digital Workerism” by co-building tools with workers that support their goals in collective organizing while preserving their autonomy, privacy, and wellbeing \cite{calacci_2022,calacci_2023}. While past HCI scholarship has mainly focused on the resistance tactics employed by gig workers, our paper examines the countermeasures used by workers across multiple types of employment, including operations, customer service, marketing, and food and beverage service.

\section{Methods}
\subsection{Content Analysis of Reddit Posts}
\subsubsection{Reddit as a Site of Study}
In line with other studies that focus on sensitive topics such as mental health, addiction, and pregnancy loss, we chose to study forums on Reddit due to its support for anonymous disclosure \cite{ammari_2018,Andalibi_2016,Andalibi_2018,van_der_Nagel_Frith_2015,gauthier_2022}, which allows people to discuss about sensitive or stigmatized topics without having to disclose their real identity \cite{Newman_2011,van_der_Nagel_Frith_2015}. Researchers also have looked to Reddit to understand the experiences of gig workers, since online communities have been crucial for gig workers to share knowledge and build community despite the distributed nature of their work \cite{sannon_2022,alvarez_2021,ramesh_2023}. In particular, Sannon et al. \cite{sannon_2022} used Reddit to understand gig workers’ experiences regarding privacy and surveillance. Expanding upon this work, we chose Reddit as a site to understand workers’ experiences with surveillance technologies across different sectors and how they used the platform to share knowledge and support one another.
\subsubsection{Data Collection}
We first identified posts that discussed workplace surveillance through manual keyword searches using surveillance-related terms such as “track,” “monitor,” and “surveil,” paired with keywords like “work” and “job.” We then randomly selected 66 posts to do a preliminary round of coding to find other relevant keywords, initial codes, and identify worker-centered subreddits to include in our study. We excluded subreddits that did not primarily focus on workers, (e.g. r/Privacy, r/Technology, r/legaladvice), subreddits that focus on managers or professional roles that do surveillance-type work (e.g. r/sysadmin, r/askmanagers), as well as industry or company-specific subreddits (e.g r/doordash\_drivers, r/mturk, r/Nanny). This allowed us to understand how information was being shared cross-sectorally between workers in different industries and to get a sense of the breadth of different types of bossware and their commonalities and differences. Finally, we excluded 1 worker subreddit that had explicit policies against data collection. In total, we selected 9 subreddits, including 2 general work subreddits (r/jobs, r/work), 5 specific to remote work (r/overemployed, r/RemoteJobs, r/remotework, r/workfromhome, r/WorkOnline) and 2 that had a political lean (r/antiwork, r/WorkersStrikeBack). Based on our initial coding, we finalized our keyword list to the following: “track,” “monitor,” “surveil,” “spy,” “productivity,” “micromanage,” and “bossware.” 

Since we were interested in understanding the impacts of COVID-19 on workplace surveillance, we only included posts written between 1/1/2019 and 12/31/2022. We also excluded posts that had less than average engagement, which was calculated by identifying the median number of comments per post according to each subreddit. Based on this criteria, we created our filtered dataset through Pushshift API \cite{pushshift}
, which contained a total of 9,362 filtered posts. For each post, we extracted the keywords used, their comments, date, title, username, URL, and text. 

\begin{table}[h]
    \centering
    \begin{tabular}{|l|c|c|c|c|}
        \hline
        \textbf{Subreddit} & \textbf{Type} & \textbf{Filtered} & \textbf{Relevant} & \textbf{Coded}\\ \hline
        r/antiwork & Political & 3441 & 473 & 70\\ \hline
        r/jobs & General & 2709 & 91 & 58\\ \hline
        r/overemployed & Remote & 549 & 49 & 12\\ \hline
        r/RemoteJobs & Remote & 497 & 6 & 6\\ \hline
        r/remotework & remote & 585 & 30 & 12\\ \hline
        r/work & General & 637 & 53 & 14\\ \hline
        r/WorkersStrikeBack & Political & 33 & 8 & 8\\ \hline
        r/workfromhome & Remote & 588 & 50 & 13\\ \hline
        r/WorkOnline & Remote & 347 & 23 & 7\\ \hline
        Total && 9,362 & 783 & 200\\ \hline
    \end{tabular}
    \caption{Work Subreddits}
    \label{tab:reddit}
\end{table}

\subsubsection{Data Analysis}
Using our filtered dataset, the first two authors manually checked for relevance to workplace surveillance. For example, we omitted posts that discussed surveillance more generally but not in work-specific contexts or used keywords in a way that did not relate to surveillance (“computer monitor”). We also omitted posts that were deleted or removed. This narrowed our dataset further to 783 relevant posts.

From this dataset, we conducted a thematic analysis on 200 randomly selected posts and their respective comments using an inductive approach, first starting with our initial codes developed during our preliminary round of analysis and then using a codebook to conduct our full analysis. Higher-level codes captured the nature of posts (e.g. “Information Sharing,” “Sharing of Personal Experience,” “Information Seeking”), the nature of surveillance (“Reason for Surveillance,” “Surveillance Type,” “Modality,” “Surveillant Actor,” “Industry”) and Experiences of surveillance (“Sentiment towards Surveillance,” “Impact of Surveillance,” “Resistance Tactic”). Our initial Reddit analysis provided high-level insight into the various types of surveillance technologies in use, the range of industries implementing these technologies, workers' understanding of what motivated managers to implement these technologies, and accounts of the impacts of monitoring and surveillance on workers. It also provided a broad view of the different resistance tactics workers employ and how they use Reddit to discuss the topic of workplace surveillance. To further protect worker anonymity, we did not include usernames or other identifying information and paraphrased or slightly altered all quotes included in this paper to reduce their searchability while still maintaining their meaning. Reddit posts are referenced by an author-assigned post ID (e.g., [R1]).

\subsection{Interviews}

\subsubsection{Participant Recruitment}

To gain a more nuanced and in-depth understanding of the experiences and perspectives of workers under surveillance, we paired our content analysis with in-depth interviews with workers and managers with first-hand experiences with workplace surveillance. We recruited participants by individually reaching out to post authors included in our dataset via Reddit’s private messaging system. We also posted a recruitment ad on 3 of the subreddits with moderator permission (/r/antiwork, /r/remotework, and /r/remotejobs). In the ad, we included a link to a screener survey to determine their eligibility for the study. The survey included questions about their Reddit username, country of residence, as well as their experiences and general opinions regarding workplace surveillance.

In total, we recruited 10 participants who came from a range of different work experiences, including in customer support, hospitality, AI data work, marketing, logistics, and construction. 6 were remote workers and four participants had management experience (P2, P3, P4, P7). Participants came from the US, Canada, India, Mexico, Japan, and the Philippines. All had experiences with at least 1 type of workplace surveillance technology, including with audio and video monitoring, emotion AI, digital activity tracking, and location tracking. To protect their anonymity, participants are referred to by their participant ID (e.g., [P1]). We provide detailed information about our participants in Table 2. 

\begin{table}[h]
    \small
    \centering
    \begin{tabularx}{\linewidth}{ 
      | >{\raggedright\arraybackslash}X 
      | >{\raggedright\arraybackslash}X
      | >{\raggedright\arraybackslash}X 
      | >{\raggedright\arraybackslash}X
      | >{\raggedright\arraybackslash}X | }
   
        \hline
        \textbf{Pseudonym} & \textbf{Roles} & \textbf{Industry} & \textbf{Country} & \textbf{Type of Surveillance}\\ \hline
        P1 & Customer Support Specialist (Remote) & Finance & US & Audio/Video Monitoring, Digital Activity Tracking, Emotion AI\\ \hline
        P2 & Telephone Interviewer/Manager & Government & Canada & Audio/Video Monitoring, Digital Activity Tracking\\ \hline
        P3 & AI Chatbot Prompt and Response Evaluator (Remote), Design Engineering Manager & Technology & India & Digital Activity Tracking\\ \hline
        P4 & Bar Owner/Manager, Food Worker & Hospitality & US & Audio/Video Monitoring\\ \hline
        P5 & Marketing Coordinator (Remote) & Media & Mexico & Digital Activity Tracking\\ \hline
        P6 & Order Entry Specialist (Remote) & Retail Logistics & US & Digital Activity Tracking\\ \hline
        P7 & Operations Manager (Remote), ESL Teacher (Remote) & Construction, Education & US & Location tracking, Emotion AI\\ \hline
        P8 & Customer Support Specialist (Remote) & Food Delivery & Philippines & Audio/Video Monitoring, Digital Activity Tracking\\ \hline
        P9 & Marketing Coordinator & Retail & Japan & Digital Activity Tracking\\ \hline
        P10 & Coffee Barista, University Research Assistant & Hospitality, Education & US & Location Tracking, Digital Activity Tracking, Audio/Video Monitoring \\ \hline

    \end{tabularx}
    \caption{We interviewed ten workers who have experienced workplace surveillance across a range of industries, professions, and geographical locations.}
    \label{tab:participants}
\end{table}

\subsubsection{Data Collection and Analysis}
With IRB approval, the first author conducted interviews in English over Zoom between November 2023 to January 2024. During each session, the first author asked questions about participants’ professional backgrounds, their experiences and opinions concerning workplace surveillance, strategies they have adopted to circumvent or receive support around workplace surveillance, as well as ideas for alternatives or improvements. Interviews lasted approximately 40-90 minutes and interviewees were paid \$40 for their participation. Each session was audio-recorded with the interviewee’s permission, auto-transcribed, and then manually checked for accuracy by the research team. 

Using the same codebook from our content analysis, we qualitatively coded our interviews in an iterative manner, adding two additional higher level codes including worker-driven alternatives (e.g., “Opt Out,” “Transparency”), and current trends of surveillance (e.g., “Outsourcing,” “Lack of Disclosure”). Our interviews extended the insights of our initial Reddit analysis, including capturing the experiences of surveilled international workers at US-based companies and managers who experienced both sides of surveillance, as the surveilled and the surveillant. The interviews also provided additional context on workers' perceptions of data and privacy laws, as well as their thoughts on Reddit as a space for support. 

Drawing together memos from both the Reddit and interview study, we iteratively revisited and refined our interpretations. Across later rounds of analysis, we identified emergent themes such as the compounding impacts of worker surveillance, workers' resistance tactics, and worker-driven alternatives, which we present as findings in this paper. With each finding, we interweave selected quotes from both our Reddit data and interviews, taking a thematic approach to capture the coherence of insights pulled across both data sources. This approach helps to provide a more holistic understanding of how workers understand and respond to surveillance across sectors, while still leaving opportunity to discuss the nuances and unique contributions that arise from each data source. However, we acknowledge that this approach may not fully capture some of the more specific nuances unique to each data source. Nevertheless, we believe that the thematic structure best captures the overarching trends and challenges that shape workers' experiences with surveillance. We have included labels (e.g., [P1] for interviews and [R1] for Reddit) to clearly indicate the origin of each data point.

\subsection{Limitations}
Although Reddit has certain benefits, as stated above, it has a number of drawbacks that most likely affected our study. Reddit users are not representative of all workers. A majority of Reddit users are young, white, male, middle-class, college-educated, and based in the US. \cite{barthel_2016}. As a result, it is likely that we have a lack of perspectives from workers who are women, minorities, older, and those who live outside the US. This also most likely omitted certain occupations and industries from our analysis, especially more feminized forms of labor, such as care work and domestic work. In addition, by focusing on general worker subreddits, we may have missed specific nuances in experiences and strategies that are tied to a specific type of job, workplace, or platform. Additionally, our choice in keywords and the use of the term “surveillance” in the recruitment ad and during interviews may encourage participation from workers who had more negative views on monitoring. However, this was an explicit choice, as we were interested in understanding workers experiences with surveillance and their resistance tactics.

Lastly, we want to acknowledge the ethical issues around studying Reddit. While we adhered to the ethical standards of our institution's IRB for our interview procedures, public Reddit data is not considered human subjects research according to US federal law and therefore not under the purview of our local IRB \cite{hhs_45_CFR_46}. Social media posters are typically not aware that their data can be used for research purposes without their explicit consent \cite{Fiesler_2018}. One way to mitigate this is to share the research output to the original community or poster. When we asked for permission to post our recruitment ad on subreddits, we let moderators know that we intended to share our findings with their community, and several moderators expressed interest in our project. However, reaching out to the original poster is more difficult. Due to the sensitivity of the issue of workplace surveillance, most posts are likely to be from “throwaway” accounts, intended to be used only temporarily before being abandoned \cite{ammari_2019}. In recruiting participants, we only reached out to a small group of  individuals at a time, prioritizing the ones we felt were the most relevant to the purpose and scope of our research. We conducted our outreach  with transparency and sensitivity to avoid undue pressure.  Although we implemented strategies to protect posters’ anonymity through paraphrasing quotes and not disclosing usernames, a posters’ anonymity cannot be guaranteed \cite{Bruckman_2002, markham_2012}.

\section{Findings}
In what follows, we outline how WSTs create a culture of distrust, stress, and reduced productivity in the workplace, prompting some employees to employ a number of resistance tactics against their use. While there is a desire for stronger governmental and organizational policies to protect workers from intrusive forms of monitoring, workers noted that a fundamental cultural shift prioritizing worker privacy and well-being is needed to drive meaningful change.

\subsection{Continuous and Opaque Surveillance Leads to Compounding Anxieties}

While some workers on Reddit and in interviews believed that some monitoring can be beneficial for accountability and training purposes, many were concerned that the level of surveillance they experienced often went beyond what they believed was necessary for their jobs. Employees were often left in the dark concerning how these systems worked, what data they collected, and how this data was used which led to multiple anxieties about privacy and lack of trust in the workplace. Rather than improve performance and efficiency, there were multiple testimonials that described how WSTs actually hindered workers’ productivity due to the immense stress they caused when these tools were used for continuous surveillance, control, and disciplinary action. Managers also expressed various anxieties about the addictive nature of surveillance and the lack of confidence it cultivated between themselves and their employees. The compounding anxieties felt by both managers and workers contributed to a culture of distrust that ultimately harmed both.

\subsubsection{Privacy Concerns and Lack of Transparency}
P1, who had been working as a remote-based customer service representative since the beginning of the COVID-19 pandemic, described the various types of WSTs she had been exposed to, including call, video, and screen monitoring, keyloggers, and AI-based sentiment analysis. \textit{“It's gotten progressively worse, I think. Jobs have been implementing more and more stricter guidelines and stricter policies, and just more and more surveillance. Like a lot more… it's been crazy, honestly.”} 

The first few companies she worked for only collected metrics that were standard in the industry, such as call times, wait times, and customer satisfaction. \textit{“That was normal…there was no sort of extra surveillance,”} she assured. However, with each subsequent new role, she was subjected to more egregious types of surveillance. One employer had a policy where new workers had to be on camera for the entire workday. \textit{“They started making you be on camera for the whole training, eight hours a day. Constantly on camera. No breaks, just lunch. And if you got off your camera, they would write you up.”} In her last role, she found out that her employer could remote into her computer and phone calls at any point in time without warning. \textit{“I actually left the job because of the surveillance part of it. I just couldn't handle it…they'd be like, ‘Oh, we're right here.’ I just thought that that was a level of surveillance that I just was not comfortable with.” } Although P1 understood the reason behind call monitoring, especially when used for training and liability purposes, she strongly disagreed with the use of other types of surveillance such as cameras and keyloggers due to privacy concerns. As someone who worked from home, she was aware that she had little ability to protect the private aspects of her home life from being captured on camera while she worked.

Across Reddit and in our interviews, multiple remote workers who were being monitored with activity tracking software such as Hubstaff and ActivTrak reported having multiple privacy concerns, especially when these software continuously tracked their computer screens, mouse movements, keyboard activity, idle time, and app usage \cite{activtrak2024, hubstaff2024}. R9 called these systems \textit{“invasive”} due to the level of tracking they were subjected to, adding that the awareness of being constantly watched and having personal data continuously collected about them made them want to resign from their job. P3, a full-time senior-level design engineer based in India who worked as a remote-based data worker training AI chatbots on the side, was especially worried about the screenshot feature of Hubstaff. Knowing that his supervisor could view and take screenshots at any point of time, he was concerned about exposing his personal data on the platform:

\begin{quote}
“What if…I had opened some private file and that could be easily taken? [...] Whatever screenshots are taken, how long they will be kept? And the URLs that I visit? Now, they say that they can see the URLs, but can they also check...my browser history? Can they check my cookies? Those thoughts are always there." --- P3 
\end{quote}

P5, a marketing coordinator based in Mexico but working remotely for a US-based marketing agency, had similar concerns, especially since the tracking software was downloaded to her personal computer \cite{smith2017can}. She said she would feel differently if her company provided her with a work computer that she could use exclusively for work purposes rather than forcing her to use her personal device. She could not afford to buy another computer on her own. She went on to explain that not only is her personal data exposed to the company she works for, but there was a possibility that the data could be used by other companies \cite{keegan_2021}:

\begin{quote}
“If they have the ability to track stuff like how many times I'm clicking on something, or what kind of websites I visit daily, I wonder what other type of things that they store [...] I still sometimes think about if they are the reason why sometimes I get like targeted ads on things… if they somehow create a profile of myself and the websites I visited, and they know that I'm a social media marketing coordinator who lives in Mexico City. There are so many things that they could find in my personal stuff to sell in order to get targeted ads, and I just don't know how safe that is.”
--- P5
\end{quote}

In their testimonials above, both P3 and P5 expressed numerous uncertainties about how they were being surveilled at work, indicating the lack of transparency and limited amount of information workers have about the nature of WSTs at their jobs. This seemed to be an explicit decision on the employers’ part. When asked whether she informed her workers about monitoring their behavior, P7 --- a remote-based operations manager at a construction company who used GPS tracking to oversee employee use of company vehicles --- explained that she did not disclose this information because she believed it would increase their anxiety. There also were managers that would make workers aware about some type of monitoring that was happening, but who would not offer the full picture. P2, who had experience working for the Canadian government as both an employee and manager, noted that sometimes the managers would notify workers that they were being monitored, but admitted that \textit{“that was maybe not followed all the time.”} 

P3 was told by his direct manager that Hubstaff would only track his website activity and take screenshots every 10 minutes and that she did not have access to the screenshots herself, but that they would be reviewed by an upper-level manager. However, P3 had suspicion that not only did his direct manager have access to these screenshots, but that Hubstaff was collecting more information than was initially disclosed. Through watching YouTube videos, he found out that HubStaff had the ability to collect data on his mouse movements, keyboard activity, idle time, and the apps he used. Additionally, he learned only through talking with his coworkers who had previously been disciplined that he must maintain a level of 85\% productivity at all times or else risk being removed from the project. However, he was still unsure about how this score was calculated. While managers were reluctant to disclose information about surveillance to avoid making workers feel anxious, our testimonials show how the lack of transparency itself led workers to feel anxiety around their lack of privacy and not knowing the full extent of how these systems worked and how they were being used by their employers. This ultimately resulted in workers feeling more suspicious and distrustful of their employers.

\subsubsection{Discipline and Punishment of Employees}
Workers described how activity monitoring heavily policed them in ways that some classified as \textit{“shaming”} and \textit{“psychological manipulation”} [R134].  P3 likened the feeling to \textit{“holding a gun to your head and [saying], ‘you have to do this like this.’”} This inevitably created a culture of fear where even seemingly minor infractions such as taking a short 5-minute break or checking social media could lead to severe consequences, such as job loss, missed pay, or disciplinary action. Workers reported getting constantly flagged by their monitoring systems for various “unacceptable” behavior. As R133 put it: \textit{“Take a second longer than expected? Penalized. Stop for a quick drink of water so you can keep working? Penalized. Make a mistake and have to stop what you're doing to correct it? Penalized.}” R134, who worked in manufacturing, posted a picture on Reddit of their monitoring software interface showing a bright red screen, explaining that if their task took them a couple seconds longer than expected or if they took a 10-second break, the screen would turn red, putting them at risk of being reprimanded. Additionally, this worker said that their workplace made their productivity scores available for everyone to see, which created a dynamic of competitiveness among workers.

Workers also reported instances when WSTs erroneously flagged them for certain behavior. P7, who previously worked as a remote-based English as a Second Language teacher, reported being subjected to emotion recognition software that would assess her bodily movements and facial expressions. Despite these systems being notoriously inaccurate, P7 described how getting flagged for “poor performance” by these systems had the additional impact of affecting her pay: 

\begin{quote}
“It would give you notes like ‘At minute three, we noticed inappropriate hand movements. There’s too much hand movements and we noticed you weren't smiling as much as you could…we only noticed smiles for 17 of the 30 minutes. And your voice inflections weren't enough.’ It was really crazy [...] And they say that your eye movements need to be more consistent, you need to be looking at the cameras straight on  [...] I don't think they were accurate judges of my classes because I used a lot of props, and I smiled a ton…even if I looked away to like get a prop or something that was counted as lack of eye contact. So yeah, I think a lot of it was really unfair and inaccurate.” --- P7
\end{quote}

P7 added that there was a three-strikes policy where they were subjected to disciplinary action, missed pay, and even the threat of termination for being flagged by the system multiple times. P1 had a similar experience with AI-based “Quality Assurance” systems that performed sentiment analysis on her calls and wrongly assessed her as being angry at customers based on her voice: 

\begin{quote}
“They would flag you for all these things…that would have your call reviewed by a supervisor and it was just completely wrong [...] AIs are not good enough to tell me what I'm doing wrong with my phone calls [...] I don't ever get mad at customers. I'm not ever gonna yell at them. And it had flagged me…three times on one call…They were doing what it said and reviewing the calls, and then that leads to more surveillance. And it's just a big snowball effect.” --- P1
\end{quote}

In both cases, AI audio and video analyses flagged workers for poor performance based on assumptions that the workers felt were disconnected from reality. Their managers seemed to place more trust in the systems than in their employees and relied on the automated reports to determine whether more active surveillance was required or whether an employee should be let go. P2 stated that managers targeted specific workers more frequently when they have a personal vendetta against them: \textit{“If someone had an inkling that someone wasn't doing something right or maybe even had a grudge, I think people were being monitored more than they should have been.”}

There were multiple reports of workers forgoing breaks to go to the bathroom or to get a drink of water due to the risk of not meeting productivity standards set by WSTs. Every time P5 needed to go to the bathroom or leave her computer even for a few minutes, she would pause the timer to make sure she was not \textit{“wasting the company[’s] time.”} However, she soon learned that she was caught in a predicament—take a break and end up working longer hours, or avoid breaks all together to finish earlier but experience intense fatigue:

\begin{quote}
“It was very exhausting for me at some point, when I noticed like geez, I cannot even take a break and go to the bathroom because that means that I'm wasting time from work. And I'm gonna have to extend my work hours, or even with lunch, I cannot…take my hour break for lunch because that means that I'm wasting a whole hour that I could use to work and finish my hours earlier… I felt so tired from that dynamic.”
 --- P5
\end{quote}

P5 described that there is a common narrative that workers have more flexibility and autonomy with remote work. However, because P5’s experience was one of being constantly tracked to ensure that she was being productive at all times, she argued that remote work is more draining than a “typical” in-person job. The constant monitoring forced workers to continuously be active all the time, even when they had nothing to do. P5 described how her coworker confessed to clicking randomly on the screen to make it appear to the app that she is being productive, even after finishing all her work. \textit{“It’s so senseless, but that’s what some of us do sometimes,”} she explained.  She went on to describe the immense physical and emotional toll this type of tracking brought, on top of the isolating nature of remote work:

\begin{quote}
“I do feel that I have to be productive every single minute. And that has affected my sleep schedule, that has affected my work-life balance in a way. And emotionally, it's just draining. I feel forced in a way to be more productive than...my peers who have normal jobs...and all that while I'm isolated and working by myself at home [...] In terms of psychological effects, I do think that it takes a toll on remote workers, because it makes you feel like you're not of trust, it makes you feel isolated from your team. And it...makes you feel like you're a robot that needs to be tracked every single time for everything they do as if your work is just a machine behind the computer.” --- P5
\end{quote}

Feelings of intense exhaustion were echoed by many workers on Reddit and throughout our interviews, and some described how overwork led to other health issues, such as depression, anxiety, and panic attacks. Many ended up being pushed to a point where they were no longer able to work and were either let go or forced to quit their jobs. The few who were able to get disability accommodations were still harassed by their managers for taking required breaks. P2 stated that monitoring systems disproportionately impacted people with disabilities because they specifically target outliers outside of the standardized “norms”:
\textit{“That would often result in invasive lines of questioning for people. Like, if someone had to go to the bathroom fairly regularly because they had IBS, or someone was dealing with a mental health crisis...you're kind of attacking them for doing this thing that they need to do.”} Further, when P2 requested a disability accommodation that would allow them to take more breaks, they were demoted and their contract did not get renewed: 

\begin{quote}
You can absolutely receive accommodations from this organization, but as soon as you inconvenience them, since you're on a short-term contract, you will just not be renewed. It's happened to so many people, it happened to me eventually, I was demoted. Because I asked for an accommodation. I'd never had a poor work review. But since I asked for this accommodation, they put me down. They said, it'll be easier for me. So I just put up to that...The laws are in place, but they have it structured to circumvent them in the first place.” --- P2
\end{quote}

As these testimonials demonstrate,  pervasive workplace monitoring created a culture of fear among workers as they faced severe consequences for minor and even false infractions, leading many of them to experience high emotional and physical exhaustion. This pressure prompted some to adopt subtle forms of resistance, while others chose to ultimately leave their jobs altogether.

\subsubsection{Uneasy Relationships in the Workplace}
WSTs created a culture of resentment and distrust amongst workers and managers when used as a disciplinary tool. 
On Reddit, workers reported having negative perceptions of their managers who heavily relied on WSTs, frequently describing them as \textit{“micromanagers”} [R6, R53, R77, P82]. P2 stated that the surveillance did not go away once they were promoted to being a manager. Initially happy about getting promoted because they wanted  \textit{“get out from under [the surveillance],”} they soon learned that they still were being surveilled by their upper-level managers with the same tracking software, although they did not have much insight into how they were being assessed.

They felt guilty about being in a position where they were doing the same surveillance they were previously subjected to.  \textit{“Since it was so bureaucratic...you had to follow things a certain way, even if you had reservations about it.”} Their promotion led them to face distrust from their former peers. They explained that even though their role technically was qualified to be part of the bargaining unit of their union, other union members would not trust them, even though they believed that their role could be an asset due to their “insider” role as a manager. Although P2 stated that they did not want to be a \textit{“cop”} and avoided disciplining workers for infractions found through surveillance technologies, they were still required to report it to their higher-level managers. P1 echoed that her managers were also monitored to make sure they were enforcing surveillance policies despite their negative feelings about it. \textit{“Everything they did was surveilled too…just like us [...] The trainers…you could tell the ones that didn't agree with it because they were being forced to tell us to turn our cameras on.”}

As a restaurant manager, P4 noted that the convenience of viewing camera footage via his mobile app anytime and any place made him \textit{“addicted”} to monitoring, which also made him more paranoid and distrustful of his workers. These feelings grew the few times he did catch workers doing something they were not supposed to: 

\begin{quote}
“I was always just paranoid about sales, so I'd always look and see how busy we are too [...] I was like, why am I watching this? Because then I'd look at it and I would just happen to see someone standing there totally out of context of whatever the actual situation is, and just assume that they're being lazy or something. So it started to definitely impact how I saw the staff. So I tried not to look at it. But I'm sure that I'm not alone in watching cameras and starting to make assumptions about what's happening in the workplace [...] It just made me paranoid…just always looking at the cameras because they were there…they just got into my brain in an addicting sort of way and definitely led me to assume kind of the worst about a situation [...] It just was one of those things I went to when I opened my phone." --- P4
\end{quote}

As these testimonials show, the nature of WSTs creates anxieties for both workers and managers and leads to poorer performance and increased disciplinary action and distrust. These systems, which are supposed to objectively monitor and improve performance instead create a feedback loop of harm for workers and managers.

\subsection{\textbf{Resistance Tactics}}

In response to oppressive WSTs, workers used various resistance tactics to regain their autonomy. Alongside commiseration, which fostered community and shared learning, they also developed individualized strategies, often inspired by these collective exchanges. These tactics included researching WSTs, using obfuscation or disabling surveillance, deliberately slowing their work pace (soldiering), and, in some cases, quitting.

\subsubsection{Commiseration}

One of the most common and accessible tactics employed by workers who were facing surveillance was simply commiserating with other workers, both online or in person. P3, for example, said that he was in regular communication with some of his colleagues through video calls, where they discussed strategies or other information, which project managers were not privy to. It was through talking to her coworkers that P7 learned tactics like mouse jiggling. Since activity-tracking apps typically measured productivity through tracking mouse movements and keyboard activity, some workers learned that they could bypass being flagged for inactivity by continuously moving their mousepad.

P5 similarly commiserated with her coworkers over a WhatsApp group. These platforms, which are separate from their company’s platforms, provided her and her coworkers with a safer space to talk about work issues. While she appreciated this support, P5 still felt isolated since many of her coworkers did not seem as concerned about the issue of surveillance:
\textit{“I don't think that they feel so strongly about this as I do. Because I voiced my concern many times, and they just keep saying, ‘Well, it is what it is, this is what we signed up for.’ [But] I don't think that this is what we signed up for, you know?”} Although P5’s friends have suggested that she talk directly to her manager about intrusive monitoring, she feared that she would face disciplinary action if she does:

\begin{quote}
“The advice that I get more and more often from my peers and from people around me, it's like, why don't you ask your boss to just remove these from your work? Why don't you ask him If it's okay that you work without your timer, and instead they give you like a base pay wage? [...] I haven't really had this conversation with my boss either. Because I feel scared that…there's gonna be pushback from that or that I might get fired or something like that.” --- P5
\end{quote}

This led P5 to turn to a remote work subreddit for advice. When P5 posted about her experience with workplace surveillance, many posters wrote that she did not deserve this treatment and should quit her job. P5 noted that this advice was completely different from the advice she received from her coworkers. This experience of sharing her story on Reddit made P5 realize that she was being mistreated. She started looking for other jobs, and because of her experiences, she is more mindful of what she values in a job, like benefits and no time tracking.

\begin{quote}
“I finally understood that I was downplaying a lot [of] what the company's doing to us, and how they're treating their employees. And I was really downplaying it, because nobody around me could tell me that, and they didn't really understand the context of what was going on. And seeing how other remote workers don't have to deal at all with that issue was eye-opening for me.” --- P5
\end{quote}

Taking learnings from the advice she received from Reddit, P5 went back to her peers on WhatsApp to encourage them to take breaks when they needed to without worrying about the time tracker: \textit{“I think I'm an instigator. But I kind of told them, ‘guys don't feel guilty if you have to stand up and go to the bathroom and take a little longer than usual because honestly, what's five minutes that you are not clicking through anything?”} Similarly, P9 commiserated with other coworkers in person about monitoring but faced a similar response. Those close to her were empathetic but told her to deal with it. Feeling isolated, she found the “antiwork” subreddit because she wanted to find a community with people who had similar work experiences. She mostly browsed but also posted from time to time to show support and solidarity for other workers who are going through similar experiences: \textit{“Being able to feel supported on this kind of big issue that was affecting me did make me feel like I wasn't alone [...] it was really helpful to know, just that I wasn't by myself.”} Being a part of the antiwork community also changed how P9 navigated her job: 

\begin{quote}
“When I first joined antiwork, I wasn't familiar with sort of the core ideology [...] Back in the day, when I had work to do, I'll do it all. I was quite obedient. But I guess after being on the sub for a while, and after having my personal experiences, I sort of developed the ideology of, don't show your work at the maximum that you can do, because it's never gonna get you anywhere good. You're not going to get a raise your hard work is going to be exploited. So just do what you need to do to stay out of trouble. But don't give them your everything.” --- P9
\end{quote}

For both P5 and P9, as well as many other workers, Reddit provided an accessible entry point for them to start questioning the presence of WSTs and discover tactics of resistance while receiving advice, resources, and support from other workers in an anonymous way. This experience led both of them to face a shift in \textit{“ideology,”} as P9 put it, that recognized the agency they had in shaping their working conditions and the wider sociopolitical aspects of their labor. Through commiseration, P5 was able to further transfer the knowledge she learned from Reddit to her coworkers on WhatsApp. By sharing experiences and strategies on these platforms, workers were able to foster a sense of community that helped mitigate feelings of isolation. This empowered individuals to start questioning and resisting against surveillance practices, while bringing others into the fold.

\subsubsection{Research}
Doing research was another accessible technique that workers employed. Some workers tried to increase their awareness of workplace surveillance by reading news articles and doing online searches about the topic. Workers on Reddit regularly posted news articles about workplace surveillance, such as to share the latest technologies in the space [R6], as well as companies that were actively surveilling their workers [R105, 192]. P1 reported reading news articles about how companies surveil workers without disclosing it and now assumed that most companies practice it. P3 used YouTube and other sources to learn more about Hubstaff and its capabilities. Through his research, P3 learned that his supervisor had the ability to take screenshots of his screen at any time without notifying him. Both P3 and P5 also tried to look into studies about workplace surveillance, but P5 acknowledged that she found little information available, especially regarding what her rights were as a worker:

\begin{quote}
“I looked for so many places, so many websites, I look for studies for research for people discussing this, and I didn't find anything, or I found very little information about it. So I would love to be part of that discussion or kind of also advocate for what I think is right, because it's so unfair. And I keep saying it’s so unfair that we are expected to comply with this software, without understanding our rights.” --- P5
\end{quote}

Through their own research, workers gained a better understanding of how surveillance technologies are operationalized. However, for some, there remained a concerning lack of information, highlighting the need for more accessible information made available to workers on the details and parameters of WST applications.

\subsubsection{Obfuscation and Disablement}
When they were actively being surveilled, some workers employed obfuscation or avoidance techniques to circumvent surveillance. P4 recalled that at his first job at a chain restaurant, he identified where the cameras were and purposefully tried to avoid them. \textit{“You can see where the dead spots are. So you can go outside, or go into the bathroom, or the walk-in cooler. These are all places you could not be filmed and do whatever.”} He remarked that he did this during times when he wanted to take a break or didn’t want to be seen just standing around even when all his work was done for the day.

Remote workers also adopted strategies to keep their work and personal files separate, such as through virtual machines or separate devices in order to keep their personal information and activities outside the purview of management. P3, for example, stated that he kept his social media and banking apps on his mobile phone. While some workers had access to multiple computers that allowed them to keep work and personal files separate, P5, who was not given a separate work computer by her company, had to resort to using her personal computer. One way she kept personal and work files separate was to create a separate user account exclusively for work-related purposes on her personal computer. However, P5 acknowledged that sometimes it is difficult to separate the two: \textit{“Sometimes I do find that some things collide, just because, while I'm working, I have to do personal tasks. But that has helped me a lot in just mentally separating those things, and not mixing my work so much with my life-related duties.”}

When unable to avoid surveillance, some remote workers like P7 employed technological hacks such as mouse jigglers to simulate activity. On Reddit, workers shared multiple variations of the mouse jiggler, including physical devices you could buy online, script-based versions, and ones constructed with everyday objects. For example, R196  shared a video of their homemade mouse jiggler made of a fan, steel bar, a Tupperware container, duct tape, and a rock, while R197  shared a photo of a bobby pin jammed into a keyboard. P9 reported using a number of apps that simulated the look and feel of common work applications that allowed her to browse the web or read Reddit discreetly during her downtime: \textit{“Each thread pops up looking like an email. So you're just scrolling along, it looks just like you're answering your emails.” } Both P7 and P9 said that the obfuscation strategies they employed were successful and they had yet to face disciplinary action or get caught.

Workers also disabled technologies while on the job. For example, P1 would sometimes refuse to turn her camera on, especially on days when she was not feeling well. However, she was careful not to make it an everyday occurrence, since she would be at risk of termination and saw coworkers fired for refusing to enable theirs: \textit{“I have told them that my camera wasn't working on multiple occasions [...] I've shut that little camera door and told them my camera wasn't working. I've told them that I lost my webcam when I moved.”} However, P1 noted that certain surveillance technologies like keyloggers were harder to bypass and many workers did not have the ability to disable monitoring systems. P5 recounted the moment she realized her tracker could not be disabled, stating, \textit{“That's when I realized, Oh, this is not optional. I just have to deal with this. And I have to suck it up if I want to get paid.”} Although some workers were able to successfully avoid, trick, and even stop the surveillance they were subjected to without detection, in a heavily surveilled workplace, there was often little room for such maneuvers, forcing workers to explore alternative strategies.

\subsubsection{Ignoring and Soldiering}
In cases where obfuscation and disablement of workplace monitoring were not possible, some workers chose to simply ignore the tracking software and instead alter how they worked. P5, for example, stated that she began to be more flexible with the timer when she realized that pausing it every time she took a break led her to work longer because she had to make up for lost time. “\textit{Every time that I pause the timer, and I'm not tracking time, that means that I have to extend my work hours.” }Additionally, she reasoned that she did not want to give her whole self to a job that didn’t value her. P5 explained that she does not receive any benefits from the job, time off, or bonuses, leaving her with few incentives to work at 100\% productivity.

Workers also learned to purposefully slow down their productivity – also known as “soldiering” – sometimes making it seem like they were being productive longer than they actually were. For  example, P2 described scenarios where they would call numbers that would be out of service or click around aimlessly at a case file. \textit{“I think the big thing that went around my office and I definitely did it too was you would kind of take your time in each case [...] I guess I was trying too hard to look like I was doing something.”}

Prior to his company implementing Hubstaff, P3 used a similar tactic with a different monitoring tool that only tracked how much time P3 spent on each prompt. This allowed P3 to finish a task but logging it as taking longer. \textit{“Imagine there's a prompt, and I could answer it within two minutes. And then probably I can slack off for another five minutes. I can work on other things. But it never used to know about it. And only when I click Submit, it used to record it.”}

P9 caught on that if she got her work done earlier, that they would give her more work to do, so she intentionally slowed her pace to avoid setting a new productivity norm:

\begin{quote}
“Once they find out that you can do more work than they have assigned you, you will get that new optimum workload. It’s kind of like a new norm, if that makes sense. And they'll expect you to always work at 110\% capacity all the time. And then because you cannot do that, your bonus will be lower because your productivity is lower, right? So no, I don't see any reason to [...] give myself extra work when I don't have to.” --- P9
\end{quote}

R134, the same manufacturing worker who previously stated that they experienced anxiety when their monitoring software showed a red screen eventually learned to ignore it, adding: \textit{“Even though I can't turn the screen off, I learned to turn the screen to the back of the machine and work at my own pace. I know this is purely manipulation, they will not fire us because we are skilled operators that work for small wages and produce lots of profits.” }Knowing that their employer may have difficulty replacing them due to their expertise gave this worker a sense of power and agency to control their pace of work in the face of constant surveillance.

Still, workers cautioned about the increased likelihood of disciplinary action should they be discovered for resisting, especially in occupations where there is heavier monitoring, such as call center work, as stated by R132: \textit{“Call centers have so many monitoring systems in place that there’s no way to escape it. You either do the job and earn your pay, or you quit.”} P2 further elaborated on this point, \textit{“The trade off is that maybe you'll get monitored while you're doing that. And since you're not on the phone, you won't notice that they're listening or looking at what you're doing. So, it would kind of take the weight off for a little bit unless they caught you.”} When disabling tactics were not feasible, workers adapted by ignoring timers, slowing their pace, and extending task times to appear constantly productive. For some, these strategies provided a sense of control over their work and time, though there remained a risk of disciplinary action.

\subsubsection{Legal Action and Collective Organizing}
Although some posters on Reddit suggested resisting through unionization or taking legal action, many of them admitted the difficulties in doing so. In cases where workers were protected through a union, some found that it did little to change their conditions when it came to monitoring. P2, who was represented by a union, noted that bringing this issue to their representatives had not led to substantial change, especially for workers on short-term contracts:

\begin{quote}
“This place would put you on very short contracts, because it's so hard to fire a government employee. If they had a problem with someone, like a legit problem, it would be a really hard process to get rid of them. But it usually would end up [with] people who definitely deserve to be there having their work cut short because of stupid things, because six months [later], they don't need to renew the contract for any reason.” --- P2
\end{quote}

Some workers on Reddit described facing strong anti-union sentiment among their coworkers, while others mentioned the trouble of living in a “right-to-work” state. As R134 put it, \textit{“if you mention unionizing, you are immediately fired.”} They added that although workers are legally protected against retaliation for unionizing, their employers \textit{“usually come up with some excuse about attendance or inefficiency and then just like that, you're out.” }Others noted the difficulties of taking legal action for employment issues. R147 added that they were \textit{“surprised by how many people suggest to just hire a lawyer and sue.”} They described the barriers their spouse experienced while attempting to seek legal counsel in a wage theft case. Because the company used a time tracking system, the poster assumed that it would be a clear-cut case. However, of the twenty employment attorneys R147 reached out to, most did not respond and the few who did told them that it wasn’t worth the time and effort:  \textit{“My partner and I are both well-educated and earn reasonable incomes, yet we've found it extremely challenging to access the legal system. I can only imagine how much harder it must be for others. In my experience, the legal system provides minimal protections for workers and largely serves to reinforce the oppressive system we live in.” } These testimonials illustrate the difficulties of collective organizing against workplace surveillance, underscoring the lack of meaningful  worker protections against its use and the need for alternative approaches.

\subsubsection{Quitting}
Ultimately, in cases where workers felt like they had no alternatives, many indeed chose to quit their jobs to get away
from intrusive monitoring. \textbf{Quitting was the most frequently named resistance tactic suggested by posters on
Reddit to deal with workplace surveillance.} During their interview, P1 noted the limited amount of power workers
have to combat surveillance, faced with only two real choices: \textit{“You can either quit, or you can stay
there and try to get a paycheck and be uncomfortable.”} However, workers stated that their awareness of and experiences
with surveillance in previous workplaces informed their subsequent job searches, explicitly choosing jobs with
less invasive monitoring and more worker autonomy, and avoiding companies where this was not the case.  Together, these resistance tactics, including quitting, illustrate workers' ingenuity in seeking to gain control over their working condition, even in the face of constant surveillance.

\subsection{Worker-Driven Alternatives}

As workers reflected on their experiences with WSTs and their methods of resistance, many recognized the limited individual power they had in enacting change. Some advocated for stronger governmental and organizational policies to better account for employee privacy and well-being. However, there remained skepticism regarding the possibility of these changes occurring in a culture that often prioritizes profits over people. This tension highlights the understanding that a fundamental cultural shift is necessary in order for monitoring technologies to truly value worker autonomy and well-being.

\subsubsection{Stronger Governmental Protections}
Despite workers’ belief that they had limited power to change how WSTs are currently used, both workers and managers surprisingly shared similar visions for how workplace monitoring technologies could be improved. Many of our interviewees advocated for stronger regulation at both the governmental and institutional levels, noting that current laws and policies meant to protect workers against surveillance were insufficient. P9 argued that regulation is needed to protect workers because companies often prioritize profits over their employees’ wellbeing. She said that without stronger governmental regulation, companies will continue to adopt more invasive monitoring practices:

\begin{quote}
“I think the only way that companies are going to stop or slow down or anything is if it comes from the government level. Some companies will adopt procedures and policies that are better and not so invasive, but a lot of them are just going to do as much as they can do legally and that's just what it is. And so I think if it's going to change, it has to come from the government level.” --- P9
\end{quote}

P1, a US-based worker, looked to Norway as a model for how workplace surveillance should be regulated, believing that the country has implemented stronger data privacy protections as compared to the US.  In addition to enforcing the EU’s GDPR, Norway introduced the Working Environment Act which limits how employers can access their employees emails and camera surveillance \cite{norway}. Other workers called for stronger regulation that would place hard limits on monitoring, provide more transparency around its use, and allow workers to provide meaningful consent and the ability to opt-out of surveillance. More specifically, P1 and P3 called for limits on more excessive forms of monitoring, such as keyboard, browser, and mouse tracking, as this type of data can be easily misused for purposes other than tracking one’s productivity. P3 also called for limitations on how personal data is collected and stored by these technologies, advocating for data to be deleted within a year or after a project is over. Additionally, P3 wanted institutions to have clearer policies around handling worker harassment as a result of surveillance: \textit{“They can misuse this in a very bad way where they can micromanage [...] They can actually see exactly what [you’re] working on every second [...] They can just use it [for] harassment. So there should be some way of telling your manager that you should not be doing so and so things which are detrimental to an employee, that [prevents] harassment.”}

However, some workers expressed more skeptical views on regulatory approaches and pointed to limitations. P5, who worked in Mexico for a US-based company as a remote worker believed that the US did not have to adhere to labor laws specific to Mexico, which she claimed contributed to poorer working conditions and no paid time off:\textit{“The current condition that I'm in, in my current employment, and the way that they operate should be illegal [...] What measures are they taking in order to comply to the legal standards of the countries that they are hiring at?”} P8, who was based in the Philippines but served clients in the US was also skeptical of stronger regulation due to distrust of her own government since, she believed, the Philippines did not have a great track record of protecting their citizen’s privacy and security: \textit{“[In the] Philippines [...] we don't have a solid foundation regarding security [...] If you're asking if I should have [regulation], for now, no, since we don't have a solid foundation regarding that security.”} P6, a US-based worker, also did not believe that regulation would stop companies because of the belief that they worked in direct collaboration with politicians and would continue to do what they wanted, regardless of regulation: \textit{“They're never gonna not collect a certain amount of information on you, because I feel a certain way. So no, they're these corporations will have free rein to do essentially, whatever they want.”} While many workers called for stronger government intervention, others were skeptical that regulation alone could adequately curb invasive monitoring practices, especially as workplaces become more fissured and global.

\subsubsection{Stronger Organizational Policies}

Beyond regulation, interviewees wanted their workplaces to foster a culture of trust, with an understanding that workers do not have to perform at 100\% at all times and also have the ability to dictate the conditions in which they work best. R151 said that it was actually poor management practices that make these tracking systems so popular: \textit{“Easily measurable factors aren’t necessarily good indicators of employee performance, while meaningful indicators are often difficult to quantify. In my view, these tracking systems stem from misguided, careless, and lazy management.”} P2 added that workers would be more driven to be productive \textit{“if they feel like they're trusted, and in a place that they're happy to work...if you're being watched at all times, it's not the best environment.”} P3, who was also a manager in his full-time role, stated that he fully trusted his employees, even during periods where they may not be as productive:
\begin{quote}
“I clearly state the goals, what needs to be achieved by what date, within what time we have to be at this milestone and all of that. So when I make these goals clearer, and if my team is following that schedule, I have no problems whether they log in two hours a day or…they slack off an entire day, they work during the entire night, I don't mind. I need to get the work done and I don't mind exactly at what time they are active, exactly at what time they are logging in and logging off. Similar is the case with my manager to me. He has given me complete freedom when it comes to me logging in and logging off.” --- P3
\end{quote}

To those we spoke with, building this culture of trust would lessen the general need for WSTs, especially those that continuously track workers by the minute. However, workers and managers acknowledged that some forms of monitoring can be helpful for training and accountability purposes. In such cases, workers wanted to have input on how surveillance is implemented and to determine what metrics are valuable to support the health of the business and workers. P6 stated that although companies do have a responsibility to improve their business, \textit{“employees should have a say in what the metrics are.”} As part of ways to use these technologies to support their well-being, workers wanted these technologies to be more accepting of accommodations. P5 felt that it should be up to the worker to determine whether to use time-tracking software: \textit{“In normal jobs, you're not expected to be productive 100\% of the time. So why remote workers should feel compelled to do that or feel forced to do that, with this kind of tracking software?”} Similarly, P3 did not believe that maintaining an 85\% productivity score daily is sustainable in the long term. Instead, employees should be evaluated by their productive averages over a longer period of time:

\begin{quote}
“There should be some slack of like, if I meet my productivity goals for one week, then I'll have at least two chances for the next week. Or there are chances when accidentally we will come get below 85\%. They may be some sudden emergency or I forget to turn off things. So there should be some way of making the employee feel comfortable about his continuity in the project.” --- P3
\end{quote}

P6, who expressed more positive views on monitoring technologies, said that it was because his company did a good job of supporting people who may fall behind on metrics. This included an extensive training period where metrics were not held against new workers. He also stated that workers were still able to manage their own schedules: \textit{“It most definitely starts from the top down, and so seeing that in practice has been a nice reassurance that they're here to work with me, instead of just get rid of me, if I don't have metrics.”} Yet, P2 questioned the possibility that WSTs could be built to support workers. They believed that these technologies can never be truly ethical unless the issue of treating humans like machines is addressed first:

\begin{quote}
“I think, if people were looked at as being human beings and not machines, that would probably be helpful off the hop, just because people do need to have some social connection during their eight-hour shifts and not be punished for it. And people do need to be able to go and take a little walk or get some water and clear their head after someone is screaming at them...just understanding that they can't just hop from case-to-case-to-case and not having the expectation in the first place. Knowing that the work will get done if they're feeling cared for and connected and have support from their workplace will probably be very productive, I think.”  --- P2
\end{quote}

Although interviewees expressed a desire for workplace cultures built on trust, flexibility, and limited monitoring, they largely found this incongruent with the use of WSTs. Their testimonials reflect a broader understanding that developing monitoring technologies that prioritize trust and worker wellbeing would first require a drastic cultural shift, fundamentally challenging the devaluation and perceived disposability of workers across the economy.

\section{Discussion}

Our findings reveal how WSTs create a culture of distrust and anxiety that negatively impacts both workers and managers. Workers reported fears of privacy invasion and disciplinary action, exacerbated by opaque and unaccommodating monitoring practices. Managers were also subjected to surveillance and experienced paranoia as a result of WSTs. Despite having limited power and legal protections, workers resisted WSTs through tactics like commiseration, obfuscation, and “soldiering,” which emphasized their refusal over compliance. Collective action, though challenging, emerged subtly through the sharing of strategies and growing political consciousness on Reddit. By drawing inspiration from these tactics, we call on the CSCW community to take an emancipatory approach to worker-centered design that supports worker autonomy, resistance, and collective action.

\subsection{The Losing Game of a Surveilled Workplace}
Throughout our interviews and in Reddit posts, workers described the multiple fears and anxieties they experienced as a result of being subjected to WSTs, even ones perceived to be less “intrusive” or“"harmful” such as productivity tracking \cite{constantinides_2022}. For remote workers, the ability for companies to surveil them in their own homes created unique privacy concerns not seen in other forms of work involving an overt and often unintended spillage of their private lives into the workplace. Workers were not only concerned about the extensive personal data being collected about them, but also that these tools were primarily used for discipline and punishment rather than support. P1 and P7’s experiences with emotion AI systems demonstrate how workers were placed under a punitive three-strikes-and-you’re-fired policy any time the system flagged them for not adhering to inflexible and unrealistic techniques of the body \cite{mauss}. As P1 noted, this created a \textit{“big snowball effect”} that made workers more susceptible to additional surveillance, suspicion, and harassment by management. 

The constant threat of disciplinary action led some workers like P5 to adopt self-policing behavior such as choosing to forego breaks when they needed them in favor of \textit{“not wasting company time.”} Foucault describes how the power of the panopticon becomes perfected when society absorbs and internalizes the established rules of the system as their own \cite{foucault}. Its ability to be both \textit{“visible and unverifiable”} allow the panopticon to establish permanence even when not being actively used by those in power. In this case, workers may be aware that they are being monitored at all times, but they were often given little information from their employers about how they were being monitored and assessed. Although some managers, like P7, reasoned that disclosing such information would make workers feel more anxious, workers responded that not having a clear understanding of the full extent of monitoring or what metrics are being used for assessment only worsened their anxieties \cite{banville2020resisting}. The compounding stress of being under constant threat of privacy invasion and disciplinary action often coalesced into feelings of extreme exhaustion and other health issues that greatly hindered workers’ performance at work. As our testimonials show, the inability of the system to accommodate and support these workers, especially those with disabilities, only intensified the punitive measures used by management and, in some cases, led to workers’ termination or resignation. 


While the harms of WSTs on workers have been well-documented \cite{awumey_2024}, we also draw attention to the harms experienced by managers. Middle managers like P2 were not immune to the surveillant gaze, as they were subjected to the same type of monitoring as their supervisees. P2’s belief that they \textit{“had to follow things a certain way”} in spite of their internal reservations about surveillance, speaks to how surveillance culture is enforced when the act of surveillance itself becomes a metric used to evaluate the surveillant’s performance. P2 described how the relationships to their coworkers shifted the moment they were promoted to manager. Although power asymmetries are innate within supervisor-supervisee dynamics, P2’s experience shows how the presence of WSTs amplified these asymmetries due to the amount of information P2 now had access to about their former coworkers. P4 also noted that the addictive and convenient nature of WSTs and the ability to check it at any time and place contributed to increasing feelings of paranoia and distrust of his workers to the point where he assumed \textit{“the worst about a situation.”} Even encountering a single infraction reinforced P4’s belief that he had to continuously surveil people on his team. 

Together, these examples show how WSTs created a reverberating culture of distrust that hurt workers and managers alike. The punitive nature of WSTs reinforced divisions within the workplace, where supervisors assumed the worst in their employees, workers felt resentment towards their bosses, and coworkers were pitted against each other. At the root of this dynamic is the dehumanization that occurs when workers and their labor are abstracted into measurable units of accountability and compared against the metrics of others \cite{Haggerty_2000}. This in turn disincentivizes workers and bosses alike from investing in the harder-to-measure interpersonal aspects of work, like building trust and open communication that are crucial to creating a healthy work environment\cite{nunez_2015}. Similar to Veena Dubal’s description of laboring under Uber's algorithmic management as \textit{“gambling”} when \textit{“the house always wins”} \cite{dubal_2023}, we echo the sentiment shared by the Reddit poster who introduced this paper — that WSTs pull workers and even their managers into \textit{“a losing game”} where nobody wins. 

According to Gould \cite{gould_2024, gould_2023}, addressing these challenges as HCI researchers requires critical examination of the measurement frameworks that underpin WSTs, as well as the assumptions that guide our understandings and critiques of these systems. We call for a shift toward worker-centered design methodologies \cite{fox2020worker}, where workers are actively involved in envisioning and designing tools and frameworks that can genuinely reflect and support the complexity and variability inherent in their work. Only by prioritizing worker voice over \textit{“stochastic machine witnesses”} \cite{gould_2024} — automated systems that reduce human activity to decontextualized metrics — can HCI researchers design systems to enhance the workplace experience such that both workers and organizations might thrive.

\subsection{The Promise and Limitations of Policy}

When it came to regulation, during our interviews, workers expressed mixed feelings. While there were some workers like P1 who believed that governmental regulation could offer workers stronger data and privacy protections, as well as prevent more intrusive forms of monitoring, others were more skeptical. Although P2, a disabled worker, received accommodations that allowed them to take more breaks under surveillance, they felt that disability laws ultimately failed to protect them from being demoted and having their contract terminated as a consequence of requesting these accommodations. Additionally, P6’s belief that \textit{“companies own the politicians”} in the United States speaks to the incredible lobbying power corporations have at all levels of government. For example, at the time of writing, the Consumer Financial Protection Bureau has introduced new guidance designed to protect workers from digital surveillance \cite{cfpb2024worker}, but is also facing increased scrutiny and calls for abolishment from powerful industry actors \cite{stempel2025tech,stratford2024delete}.

State privacy laws such as the California Consumer Privacy Act (CCPA) introduce important protections granting workers the right to know when employers are monitoring them and for what purpose, as well as the ability to access to their personal data, request its deletion, and limit its sale \cite{feng2023overview}. However, Calacci and Stein \cite{calacci_stein_2023} highlight how current data protection laws such as CCPA fail to protect workers as a whole because they focus on protecting the individual data subject, rather than protecting collective rights. While there has been growing interest at the US federal level to introduce stronger worker protections, there are currently no federal laws in place \cite{casey_2023,deluzio_2024,ostp_2023}. We must acknowledge that most meaningful policy change for workers remains a slow and uphill battle.

The experiences of international workers such as P5 and P8 shed further light on the limitations of regulation as more jobs are "offshored" to countries with cheaper labor and weaker labor protections \cite{prb_2008}. Since P5 worked remotely for a US-based company, she believed that she did not have the same labor protections as those who worked for Mexico-based companies because her employer did not have to comply with Mexican labor laws. P8 felt little hope for stronger privacy regulation in the Philippines because, in her eyes, the country lacked a \textit{“solid foundation regarding security”} in the first place. As work becomes more fissured and more US-based companies continue to use outsourced contract labor from countries like India and the Philippines for their customer support and data work \cite{tan_2023}, this brings new questions about the role of HCI in policy to protect workers across the global supply chain. While we see great promise in the role of worker-centered design to inform governmental policy at the local, state, and even federal level \cite{yang_2024}, it is certainly not a panacea.

In the face of limited regulation, there could still be opportunities for HCI to inform stronger worker protections at the organizational level. P3, for example, advocated for stronger anti-harassment policies in the workplace to prevent bosses from overly surveilling their workers. In interviews, workers expressed their desire to be in workplaces where they were trusted, valued, and had agency over their working conditions. Taking a worker-centered approach could help to inform whether and how monitoring technologies should be implemented in the workplace, what metrics are valuable to employees, more accommodating to different styles of work, and, perhaps most importantly, contestable \cite{kawakami_2023, sannon_2022}. However, we echo P2’s sentiment that the implementation of WSTs can never be truly ethical unless workers are valued as full human beings. The logics of productivity and surveillance are deeply rooted in systems far bigger than any individual employee, manager, and workplace and will thus require fundamental cultural shifts from both above and below.

\subsection{The Power of Resistance and Collective Action}

\subsubsection{Countering Surveillance through Refusal}

In our interviews and on Reddit, workers discussed the various ways they countered WSTs. For some workers like P3, these tactics were learned through commiseration with their own coworkers. However, during times when workers felt isolated and alone, many described turning to Reddit for support. For P5 and P9, worker communities on Reddit provided them with a supportive space to anonymously discuss and seek advice on WSTs. Workers shared technological hacks like mouse jigglers to simulate activity and virtual machines to keep work and personal files separate. Some discussed approaches such as “soldiering” to purposefully slow down their productivity, while others pointed to quitting as outright refusal. While the act of quitting may seemingly speak to the limited power workers have within their current workplaces, it notably remains one of the few means by which workers have full agency to better their working conditions. 

Through these everyday acts of resistance, workers creatively and ingeniously repurposed tools, subverted expectations, and developed new methods to reclaim their autonomy. Drawing from the radical traditions of Luddism and unmaking, we echo Sabie et al.’s \cite{sabie_2023} call for HCI researchers to take an emancipatory approach to worker-centered design --- one that draws inspiration from the existing approaches workers are already using to dismantle, disarm, and destabilize systems of control. This approach could involve designing tools and mechanisms that enable workers to obfuscate, disrupt, or neutralize WSTs, as well as helping workers to safely enact "soldiering" practices that evade detection. More crucially, taking an emancipatory approach would go beyond merely redesigning WSTs to mitigate harms and toward the possibility of their abandonment when they fail to support workers \cite{johnson2024fall}.

\subsubsection{Building toward Collective Action}
While formal collective action was regularly \textit{recommended} as a resistance tactic by Redditors in our dataset (e.g., union activity or work stoppages such as strikes), its relative absence as a practice \textit{enacted} by those we heard from reveals the difficulty of carrying out such measures. Nonetheless, our findings point to how individual modes of resistance may evolve into forms of collective action. For instance, P2 adopted a slower work pace to prioritize well-being after learning this tactic from coworkers. P5 shared strategies adopted from Reddit to encourage her peers to take breaks and disregard time trackers. It was by engaging with other anonymous workers on platforms like Reddit that enabled some workers like P5 and P9 to critically assess their own relationship to work and build new political consciousness. Worker communities on Reddit were central in educating members on tactics of refusal, helping them to counter the coercive self-policing behavior they previously internalized, as well as inspiring them to motivate others to take similar actions.

Given the difficulties of more deliberate forms of collective action, we argue that support for these more subtle approaches remains critical in moving toward more robust policies that truly protect workers’ privacy and dignity. While past CSCW and HCI scholarship has developed technological tools to help workers track, share, and make sense of data that can, in turn, be used for collective action \cite{calacci_2022,calacci_2023,calacci_stein_2023,sannon_2022}, we see additional opportunities for CSCW researchers to help bridge the gap between individual acts of resistance and collective action against surveillance. Taking inspiration from communities like r/antiwork as “counterpublics” \cite{ticona_2023} — or, radical spaces where workers openly and safely share complaints, build solidarity, exchange tactics, and deepen their political consciousness — we can see potential avenues for CSCW researchers to facilitate similar forms of information sharing and mutual support \cite{salehi_2015, wu2022reasonable}. 
CSCW researchers can also take a more proactive stance by creating knowledge bases and accessible resources \cite{abu2023diverse, whitey2021hci} that empower workers with information on surveillance technologies and effective resistance strategies, as well as reflective tools that enable workers to critically examine and process their own experiences. More critically, we urge CSCW researchers in this space to more directly engage with online worker communities, taking on roles as community members or accomplices \cite{asad2019academic}.

Rather than largely fixing our gaze, as a scholarly community, on the potential for redesigning surveillance platforms or shifting managerial practice, we can support these more extended forms of worker community building that scaffold political consciousness, circulate tactics of subversion, and prefigure more sustainable and trusting working relations. Unlike technical interventions that might seek to mitigate the most outsized harms caused by intrusive surveillance, this approach would mean the slow and intentional work of collective transformation and cultural change. It is only through this recommitment to meeting workers where they are and sharing in mutual learning that we can finally unmake the surveillance apparatus. 

\section{Conclusion}
In this paper, we have sought to understand the effects of the rapid expansion of workplace surveillance technologies that followed the mass move to remote work during the COVID-19 pandemic. In our analysis of anonymous posts on Reddit and of ten interviews with workers from a range of industries, we found that constant and layered surveillance created major privacy concerns for workers, deteriorated trust between supervisors and supervisees, and led to significant anxiety and, ultimately, burnout. In understanding how workers responded to and resisted these effects, we call on CSCW researchers to adopt more emancipatory approaches to worker-centered-design that center on resistance and refusal.

\section{Acknowledgements}

We would like to thank our interview participants who generously contributed their insights, as well as the Reddit moderators who expressed support for this project. We also thank our reviewers and colleagues who 
provided valuable feedback, including Hunter Akridge, Wilneida Negrón, Tucker Rae-Grant, Matthew Scherer, and Franchesca Spektor.

\bibliographystyle{ACM-Reference-Format}
\bibliography{sample-base}

\end{document}